# Loss analysis of Low Bandgap (Ag,Cu)(In,Ga)Se$_2$ Solar Cells for Tandem Applications

*Francesco Lodola*[*], *Sevan Gharabeiki, Maximilian Krause, Shiro Nishiwaki, Romain Carron and Susanne Siebentritt*[*]

Francesco Lodola, Sevan Gharabeiki, Susanne Siebentritt

Laboratory for Photovoltaics, University of Luxembourg, Physics and Materials Science Research Unit, 41 Rue du Brill, Belvaux, L-4422 Luxembourg

Maximilian Krause, Shiro Nishiwaki, Romain Carron

Laboratory for Thin Films and Photovoltaics, Empa – Swiss Federal Laboratories for Materials Science and Technology, Ueberlandstrasse 129, Duebendorf, 8600 Switzerland

*E-mail: francesco.lodola@uni.lu, susanne.siebentritt@uni.lu

**Abstract**

Tandem solar cells can better harness the energy of the solar spectrum. Chalcopyrite solar cells have drawn attention, being the only highly efficient devices with bandgap around 1.0 eV, suitable for bottom cells. In the quest for better efficiencies, we conduct a complete loss analysis of 1.0 eV bandgap (Ag,Cu)(In,Ga)Se$_2$ cells with efficiencies around 18.5%. We perform absolute photoluminescence, electroluminescence, *JV* and EQE measurements on the absorber and the finished cells to analyze losses of short-circuit current, open-circuit voltage and fill factor. The relevant losses in current are due to absorption losses in the absorber and could only be mitigated by light management structures. But the most significant losses are found in the voltage, due to non-radiative recombination in the absorber, and the fill factor, due to a high diode factor. The diode factor of the cells is significantly higher than in the absorber alone, indicating a strong influence of recombination in the space charge region.

## 1. Introduction

The Shockley-Queisser (SQ) limit defines the theoretical efficiency limit for single junction solar cells, based on the detailed balance principle.[1] Although the model is greatly simplified, it is useful to quantify how far a given photovoltaic (PV) technology deviates from the ideal limit.[2]

Nowadays, the rising energy demand and the urgent need for clean production call for the large-scale implementation of PV.[3] In particular, tandem solar cells can better harness the solar spectrum reaching a theoretical efficiency of 46%,[4] thanks to superior spectral management, using a top and a bottom cell, compared to the 33% of single junctions. Thus, the development of highly efficient bottom cells is essential. To achieve the highest efficiencies, 1.0 eV bandgap bottom cells are required[4] and chalcopyrites are currently the only high-performance technology capable of meeting this need.[5]

Cu(In,Ga)Se$_2$ (CIGS) can reach 1.0 eV bandgap with low or no gallium content. Ga introduction in the absorber is mainly to maintain a band gap gradient to reduce the non-radiative losses at the metallic back contact,[6,7] along with an improvement at the CdS/absorber interface.[8,9] Introducing Ag[10] and heavy alkalis[11,12] has led to a further increase in efficiency. Low bandgap CIGS have demonstrated power conversion efficiencies (PCEs) of 20.3%[8] and 19.2%,[12] with 1.01 eV and 1.00 eV bandgap, respectively. Moreover, a PCE of 21.2% has been achieved with a 1.03 eV bandgap CIGS on a lightweight and flexible substrate.[9] These PCEs are close the record CIGS cells, with higher Ga content.[13,14] Furthermore, promising bottom cells have also been obtained on transparent contacts,[15] for bifacial tandem cells, and on flexible substrates,[9] enabling the progress in lightweight and bendable tandem devices.

However, the PCEs of these solar cells are far from the SQ limit of 31.6%, calculated considering a perfect back reflector.[16] Here, we perform optical and electrical measurements on a high-efficiency (Ag,Cu)(In,Ga)Se$_2$ sample with 1.00 eV bandgap, similar to the record solar cell,[12] to understand the losses in the three parameters defining the solar cell efficiency: short-circuit current ($J_{SC}$), open-circuit voltage ($V_{OC}$) and fill factor ($FF$). We demonstrate that the improvement of long wavelength collection for high $J_{SC}$ remains a challenge for these bottom cells. However, $V_{OC}$ loss presents the main fundamental loss. The $V_{OC}$ loss is separated into the different mechanisms [17,18]: non-radiative losses are the principal loss. $FF$ loss is examined by optical diode factor (ODF)[6,19,20] measurements on absorbers and on cells and their comparison with the electrical diode factor (EDF) from one-diode fits of $J$-$V$ curves. We conclude that the leading cause of $FF$ loss is an increase in the EDF value when the transparent conductive oxide is placed and the p-n junction formed. This finding is common across different bottom cells, with diverse designs. It is most likely due to Shockley-Read-Hall (SRH) recombination in the space-charge region (SCR).

## 2. Discussion

The samples studied are highly efficient 1.0 eV bandgap (Ag,Cu)(In,Ga)Se$_2$ back-graded bottom cells, labelled NaF10 in a previous paper.[21] We investigate two cells next to each other on the same substrate. **Figure 1a** shows the *J-V* curves. The efficiency is around 18.5% for both cells, close to state-of-the-art record efficiency.[8,12] We can see that the three parameters deviate substantially from the SQ limit[16] (**Figure 1b**). The cells produce 89.5% of $J_{SC}$ from SQ, 77% of $V_{OC}$ and 83.9% of *FF*. **Figure 1c** displays the external quantum efficiency (EQE) of the two cells. Obviously, the devices depart from the idealized SQ solar cell (dashed line). Indeed, in SQ one of the main assumptions is that the absorptance (or in this case EQE) is a step-function equal to 100% above the bandgap and 0 below.[1,2] The bandgap of the cells is determined by the maximum of a Gaussian fit of the EQE derivative (**Figure 2d**). It is noticeable that above the bandgap energy, at wavelengths below 1240 nm, a portion of short-circuit current is lost (light grey area). This is due to reflection losses, all over the wavelength range, and to parasitic absorption in the Al:ZnO transparent conductive oxide (TCO), i-ZnO and in the CdS buffer layer, at low wavelengths. The absorption loss at low wavelengths is not considered crucial in the functioning of these cells as the bottom device in tandems, although optimization of the buffer might be needed when higher energy photons are not present.[22] The additional $J_{SC}$ loss near the absorption edge can be reduced by a longer collection length,[23] for example by improving the lifetime with more precise alkali supply, and by a TCO with less free-carrier absorption, such as bias-ZnO or InZnO$_x$.[12,24] However, as discussed in detail later, the comparison of the absorption edge of the finished cell, obtained from absolute photoluminescence (PL), is almost the same as that calculated from EQE (**Figure 5**). This indicated that collection losses, which occur mainly in the higher wavelength region close to the bandgap, do not play a major role in these highly efficient cells. The current loss near the absorption edge is rather due to an absorptance loss, which could only be mitigated by a thicker absorber region with 1.00eV band gap, i.e. before the Ga back gradient, or by light management structures. In total, all these losses account for about 5 mA cm$^{-2}$ loss in $J_{SC}$.

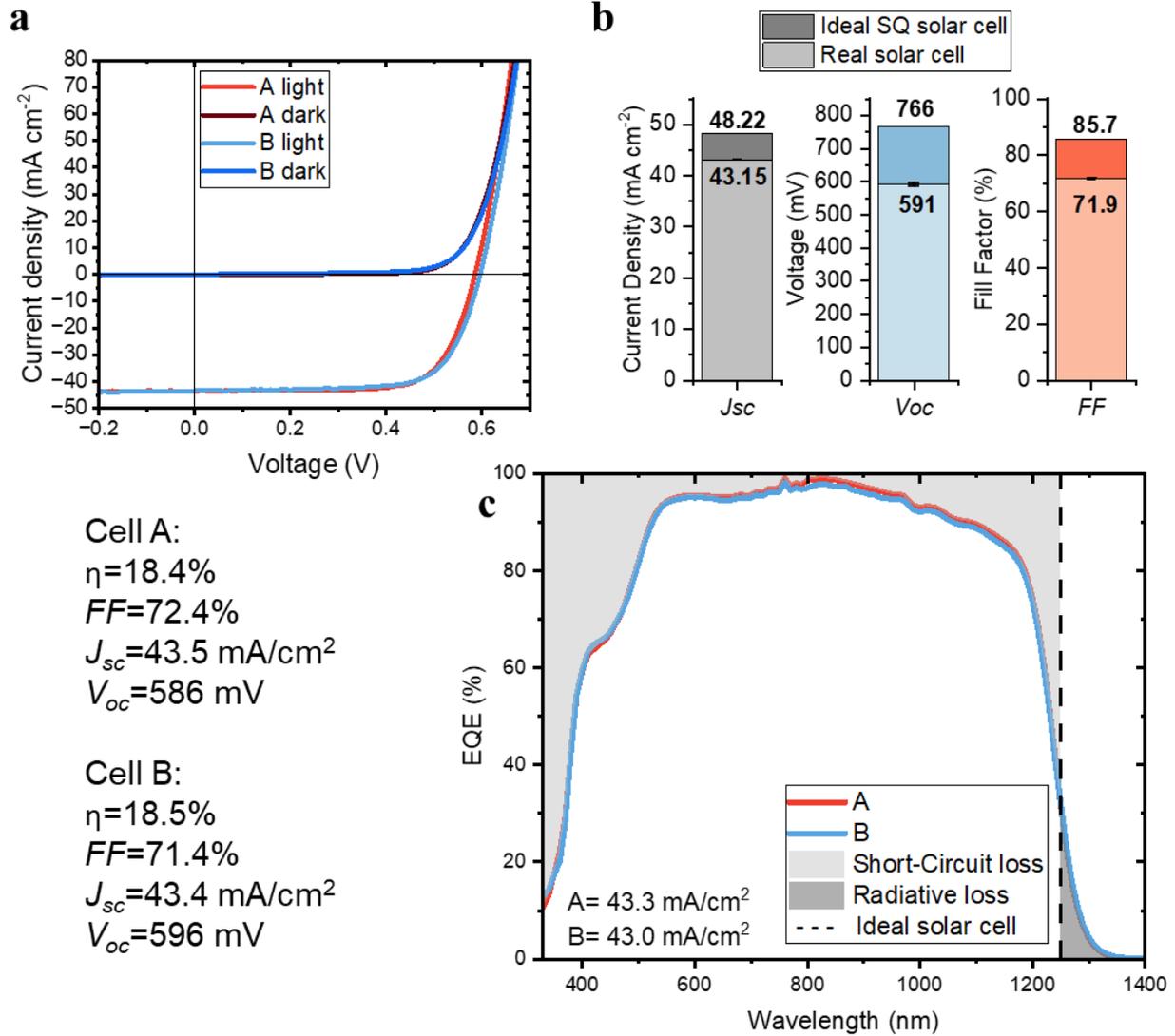

**Figure 1.** a) *J-V* curves of the studied samples; b) $J_{SC}$, $V_{OC}$ and *FF*, averaged over the two samples compared to their Shockley-Queisser (SQ) limit; c) EQE spectra of the samples, with the SQ EQE in dashed lines. The spectral part leading to short-circuit voltage loss is in light grey, while the one leading to radiative voltage loss in dark grey. The expected $J_{SC}$s obtained by integrating the EQE multiplied with solar flux spectrum, are also given for both cells.

Next, we discuss the losses in $V_{OC}$. We separate the different $V_{OC}$ loss mechanisms based on Rau's formalism[17]: short-circuit (or generation) loss,[18] radiative loss and non-radiative loss. All equations used for the analysis are given in **Table 1**. The light grey area in Figure 1c leads to a short-circuit loss in voltage of 2.8mV (**Equation 1**). The dark grey area, instead, is related to sub-bandgap absorption. It is affected by multiple factors like bandgap fluctuations[25,26] and tail states[27] and defect states. Although, this part does not decrease $J_{SC}$, it decreases the $V_{OC}$, due to an increase in black-body radiation at higher wavelengths. At $V_{OC}$ generation and recombination is balanced. To determine the radiative $V_{OC}$ loss, it is assumed that all recombination is radiative and the generation flux is balanced by the emission flux. According to Planck's generalized law[28] (**Equation 2** in Table 1) a higher black-body emission requires a lower quasi-Fermi level splitting (QFLS) to balance the generation flux.[18]

**Table 1.** Equations used in this study. All the symbols are explained in **Table 4** (Appendix).

| | Equation | | Ref. |
|---|---|---|---|
| Short-circuit loss | $\delta\Delta\mu^{sc} = -k_B T \ln\left(\dfrac{J_{SC}}{J_{SC}^{SQ}}\right)$ | (1) | [18] |
| Generalized Planck's law | $\Phi_{PL}(E) = A(E)\Phi_{BB}(E)\exp\left(\dfrac{\Delta\mu}{kT}\right)$ | (2) | [28] |
| Fit based on generalized Planck's law | $\ln\dfrac{h^3 c^2 \Phi_{PL}(E)}{2\pi E^2} = \dfrac{\Delta\mu - E}{kT}$ | (3) | [28,29] |
| Absorptance from photoluminescence | $\dfrac{\Phi_{PL}(E)}{\Phi_{bb}(E)\exp\left(\dfrac{\Delta\mu}{kT}\right)} = A(E)$ | (4) | [29] |
| Gaussian fit of derivative of A(E)/EQE(E) | $\dfrac{dA(E)\text{ or }dQE(E)}{dE} = C_0 + \dfrac{C_1}{\sigma_{(E_g)}\sqrt{2\pi}}\cdot\exp\left(-\dfrac{(E-E_g)^2}{2\sigma_{(E_g)}^2}\right)$ | (5) | [18,25] |
| Absorption coefficient from Absorptance | $\alpha(E) = -\dfrac{\ln(1-A(E))}{d}$ | (6) | Lambert-Beer's law |
| Urbach energy fit | $\alpha(E) = \alpha_o \exp\left(\dfrac{E}{E_U}\right)$ | (7) | [30] |
| Photoluminescence quantum yield | $Y_{PL} = \dfrac{\int_0^\infty \Phi_{PL}(E)\,dE}{F_{Gen}^{Sun}} \approx \dfrac{\int_0^\infty \Phi_{PL}(E)\,dE}{F_{Gen}^{Laser}}$ | (8) | [31] |
| Non-radiative losses | $\delta\Delta\mu^{nr} = -k_B T \ln(Y_{PL})$ | (9) | [17,31] |
| Radiative losses | $\delta\Delta\mu^{rad} = \dfrac{\sigma^2}{2k_B T}$ | (10) | [25] |
| Optical diode factor (ODF) | $G^{n_{opt}} = Y_{PL}$ | (11) | [32] |

**Figure 2a** shows the absolute photoluminescence[29,31] (PL), replotted according to **Equation 3** in Table 1. The PL was measured on the absorber/CdS stack, obtained by etching the window layers of the cells on the same absorber used for cells A and B, although not on those specific cells (**Figure 8** in Experimental methods). By fitting the high energy wing of the PL spectrum with Equation 3 in Table 1 [29,33], the QFLS of the absorber, which is analogous to the device $V_{OC}$[31], can be extracted. Within error, QFLS is the same as $V_{OC}$ of the finished devices. There is a small fluctuation between different spots (**Table 2**, Figure 8 and **Figure 9**). A dip is visible in the PL spectrum around 0.91 eV (red box in Figure 1a). This dip has been seen in other cells as well[29] and is attributed to the absorption of water vapor due to the non-zero humidity in the lab.[34] The knowledge of QFLS allows us to calculate with Equation 3 the absorptance A(E) in the energy range covered by the PL spectrum (**Figure 2b**). All the spots measured on the absorber show a very similar shape of A(E) (**Figure 9b**). We fit the first derivative of A(E) (**Figure 2c**) and also of QE(E) of the cells (**Figure 2d**) with a Gaussian function (**Equation 5**),[18,25] to obtain the band gap from the maximum and the width of the band gap distributions. We use **Equation 10** to determine the radiative loss. No difference in σ is noticed when top layers are added onto the absorber, as seen from the comparison between the derivative of A(E) (Figure 2c) and QE(E) (Figure 2d). Possible parasitic absorption or transport losses do not play a role here, i.e. the absorption edge broadening is a property of the absorber itself. Alloy disorders due to presence of In and Ga on the group III sites, and Cu and Ag on the group I sites, as well band gap gradients,[18] dislocations, grain boundaries, strain and electrostatic potential fluctuations[35,36] contribute to the A(E) edge broadening.[37,38] The σ value of 27.5 meV is low for CIGS samples, only comparable to the value of CIGS samples without gradient.[18] The quantities of Ag (with [Ag]/([Ag]+[Cu]) ≈ 2%)[21] suggest that alloy disorder in this case plays only a minor role in the broadening of the absorption edge, σ. The Ga gradient, introduced for passivation purposes, is a 'hockey stick'-like gallium profile with Ga only concentrated close to the molybdenum back contact.[13,21] The GGI remains below 10% within the first 2 μm of the absorber and drops below 1% within 1.5 μm.[21] Therefore, the Ga gradient in these cells does not significantly contribute to the broadening of the absorption edge. Indeed, most long-wavelength photons, which are near the absorption edge, are absorbed before reaching the Ga-rich region. As a result, the absorber behaves similarly to flat $CuInSe_2$, without Ga alloying, compatible with the previously observed broadening.[18]

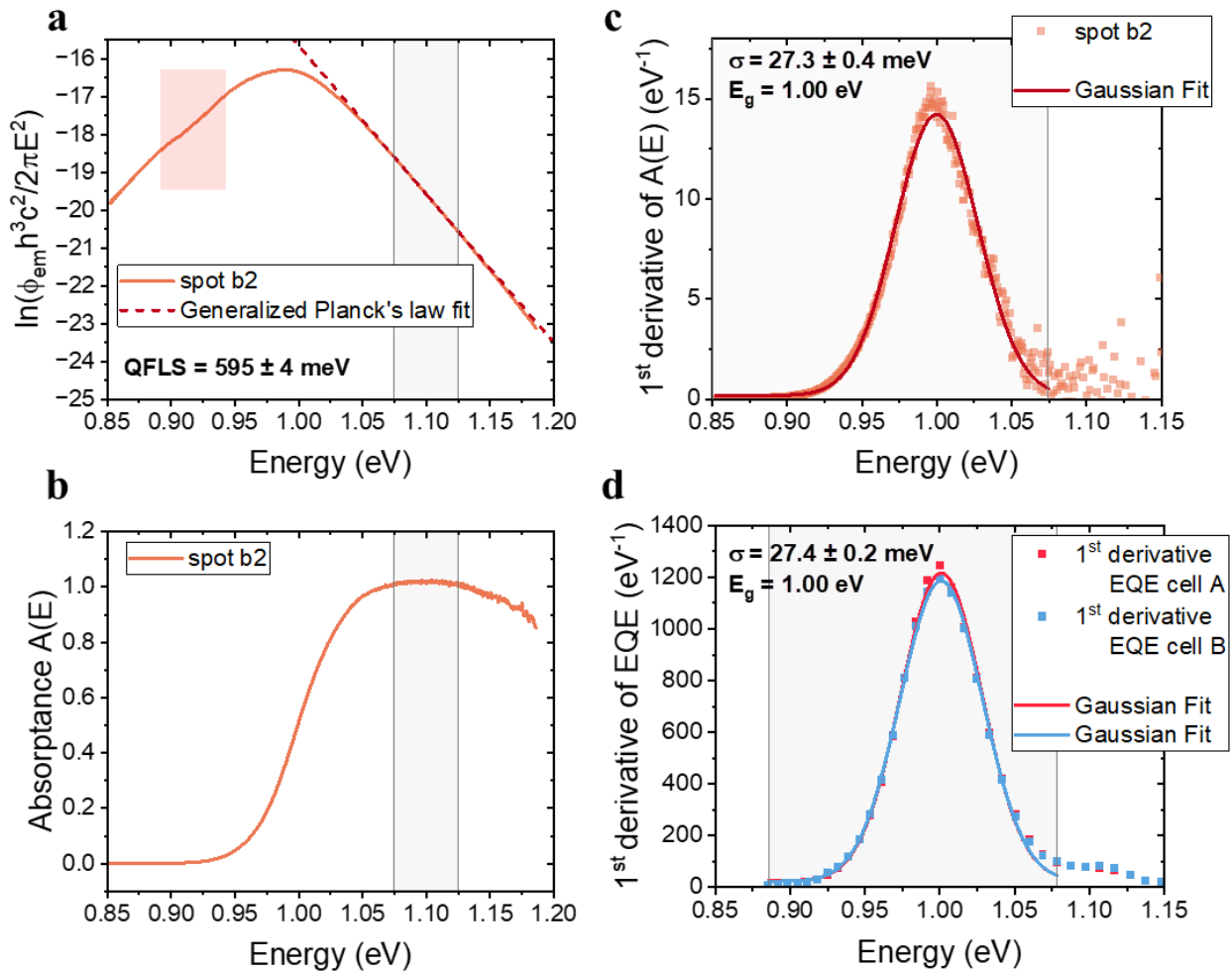

**Figure 2.** a) absolute PL spectrum, replotted with Planck's generalized law (Equation 3), the QFLS given is the average over many spots (see Table 2), of the absorber/CdS stack measured at spot b2 (for the location see Figure 8) as an example (in grey the QFLS fit area). The red box highlights the dip at lower energies attributed to the absorption of water vapor; b) PL extracted absorptance spectrum; c) first derivative of the extracted absorptance spectrum fitted with a Gaussian function; d) first derivative of the EQE spectra fitted with a Gaussian function. In black, average QFLS and σ value of all the spots/cells measured with their standard deviation.

**Figure 3a** and **Figure 3b** depicts the Urbach energies ($E_u$), which indicate static disorder, electrostatic potential fluctuations and dynamic disorder.[30,36] They are obtained from fitting the absorption coefficient using **Equation 6**. The absorption coefficient is determined from the absorptance from PL with Equation 5 or from the EQE, assuming $EQE(E) = A(E)$ near the absorption edge (see Figure 5). The lower bound for the fitting range differs between the two, as the EQE is noisier below the bandgap. In the absorption coefficient, extracted from the PL spectra with Equation 5, the fit is possible even in a region (red fitting range) where the absorption coefficient is below 10 cm$^{-1}$. In this way, the Urbach energies extracted are not affected by the absorption onset broadening, σ.[18,39] Moreover, we use an additional fitting range (grey fitting range) to ensure comparability with the EQE fit. In the end, this approach limits the reliability of the band tail fits in both EQE and in the PL absorption coefficient, since the fit is performed in a region still affected by the onset broadening, overestimating the result.[18,39] In fact, $E_u$ calculated in the range where the absorption coefficient is below 10 cm$^{-1}$ are about 0.7 meV lower than those obtained from the upper range. Nevertheless, we consider the results compatible with each other, with the finished device having slightly more pronounced band tails, as confirmed from PL on the finished cells in **Figure 4a**. This is because the onset broadenings from PL and EQE (Figure 2c and Figure 2d) are the same, so they have a comparable impact on the fits. Moreover, reported $E_u$ values of record efficiency bottom cells[8,12] are likely overestimated because the range of the EQE fit is not adequate for the estimation of $E_u$ and there is an influence of the absorptance broadening.

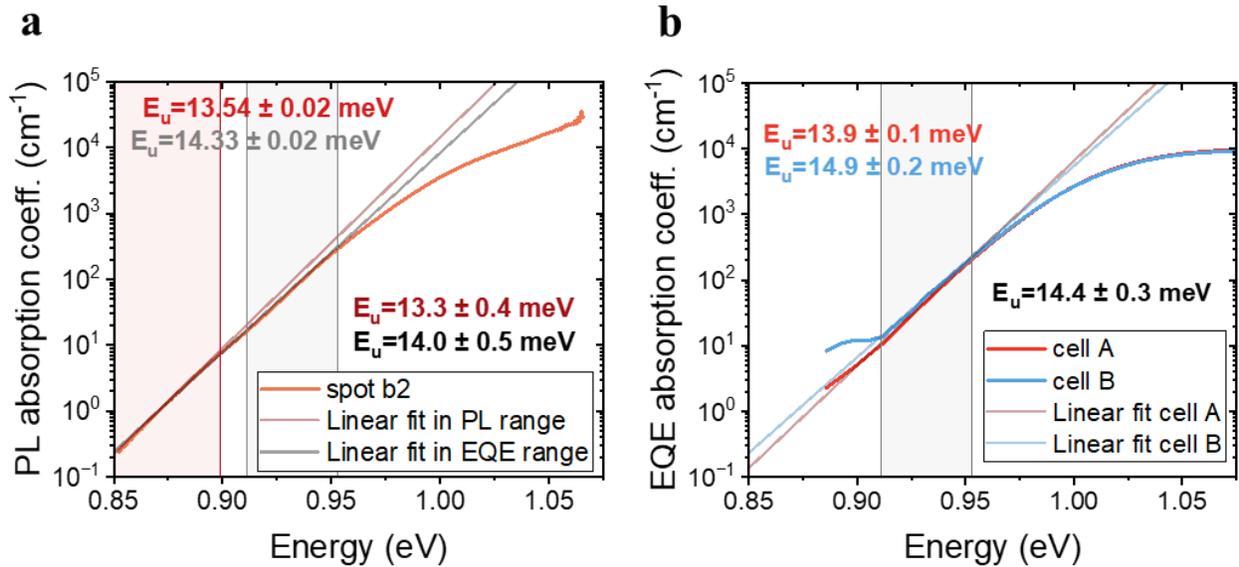

**Figure 3.** a) absorption coefficient on spot b2 from the PL extracted absorptance spectra with its Urbach energy fits from the two displayed ranges. Average $E_u$ value of all the spots measured with its standard deviation (in dark red from lower fitting range and in black from the upper range); b) Urbach energy fits of the EQE absorption coefficient spectra. Fitting ranges in grey.

Urbach energies in Table 2 scatter slightly in the same absorber (all the spots are measured on the same absorber but referred to different etched cells as explained in Experimental methods – Figure 8). It is essential to observe that PL on the absorber, in these highly efficient bottom cells, can already predict the Urbach energy values and the band gap broadening σ of the final device, in addition to $V_{OC}$ from QFLS. This implies that these parameters are limited by the quality of chalcopyrite absorber itself and not by the top layers, deposited to finish the device.

**Table 2.** QFLS, absorption broadening σ and Urbach energies extracted from PL of the various spots (see Figure 8) analyzed on the absorber/CdS stack. The displayed errors belong to the single spot measurement and are below 1 meV for QFLS and 0.1 meV for Urbach energies. Urbach energies are evaluated from Figure 3a in the red fitting range (and in the grey fitting range).

| Absorber/CdS spot | QFLS (meV) | σ (meV) | Urbach energy (meV) |
|---|---|---|---|
| a1 | 599 | 26.7 ± 0.5 | 12.7 (13.3) |
| a2 | 593 | 27.2 ± 0.2 | 13.0 (13.7) |
| b1 | 597 | 27.0 ± 0.5 | 13.6 (14.1) |
| b2 | 600 | 27.6 ± 0.2 | 13.5 (14.3) |
| b3 | 587 | 27.3 ± 0.2 | 13.1 (13.6) |
| b4 | 595 | 27.4 ± 0.2 | 13.8 (14.8) |
| c1 | 595 | 27.8 ± 0.2 | 13.4 (14.0) |
| Average ± stand. dev. | 595 ± 4 | 27.3 ± 0.4 | 13.3 ± 0.4 (14.0 ± 0.5) |

In Figure 4a absolute PL on the absorber/CdS stack (light green) and the finished cells (dark green) are measured with a red laser beam, covering the full devices as explained in Experimental methods. The QFLSs obtained are (594 ± 4) meV for the finished cells, compatible with the $V_{OCS}$ measured with the *J-V* scans (Figure 1a). We remeasure the absorber/CdS stack with the large laser spot and obtain a QFLS of (599 ± 4) meV, averaged over different detection spots (see **Table 3**), in agreement with the ones obtained with a smaller beam (Figure 2a and Table 2).

**Table 3.** QFLSs extracted with Equation 3 from the green area in **Figure 10a**. Samples excited with the large laser beam spot in the grey abc area of Figure 8. Non-radiative losses from Equation 9 of the PL spectra in Figure 10a. Urbach energies obtained from fit of the absorption coefficient

of the absorbance in **Figure 10b** with Equation7. Fitting range in grey in Figure 10a. Average and standard deviation of the spots on the same cell are also reported. Last column shows $V_{OCS}$ of the two cells measured in Figure 1a.

| Structure | QFLS (meV) | Non-radiative losses (meV) | Urbach energy (meV) | QFLS (meV) | $V_{OC}$ (mV) |
|---|---|---|---|---|---|
| Absorber/CdS a1 | 598 | 155 | 12.0 | 600 | 586 |
| Absorber/CdS a2 | 602 | 150 | 12.5 | ± 2 | (A) |
| Absorber/CdS b1 | 600 | 151 | 13.0 | 601 | 596 |
| Absorber/CdS b2 | 603 | 149 | 12.8 | ± 2 | (B) |
| Absorber/CdS b3 | 603 | 149 | 12.8 | | |
| Absorber/CdS b4 | 598 | 154 | 12.8 | | |
| Absorber/CdS b5 | 600 | 152 | 12.8 | | |
| Absorber/CdS c1 | 591 | 161 | 13.1 | 591 | |
| Finished device a1 | 586 | 170 | 12.4 | 586 | 586 (A) |
| Finished device b1 | 593 | 161 | 13.4 | 594 | 596 |
| Finished device b2 | 594 | 160 | 13.3 | ± 1 | (B) |
| Finished device b3 | 596 | 159 | 13.3 | | |
| Finished device c1 | 599 | 155 | 13.5 | 599 | |

In an ideal SQ solar cell, only radiative recombination takes place, which results in the absence of other recombination mechanisms.[1,2] However, real absorber and cells have non-radiative recombination paths.[17,31] These can be evaluated with PL, through the photoluminescence quantum yield $Y_{PL}$ (**Equation 8** and **Equation 9**)[17,31] and are given in Figure 4a. Higher non-radiative losses, about 8 meV more, are noticed for the finished device, compared to the absorber/CdS stack, visible already from the slightly lower PL emission from the finished device. Within error this loss agrees with the difference in QFLS between the absorber/CdS stack (599 meV) and the QFLS of the finished cell (594 meV). We may attribute this difference to optical losses due to parasitic absorption in the sputtered i-ZnO and AZO top layers.[12,24,40] The deposition via sputtering of these layers may also cause damage,[41] together with the interdiffusion of Cd from the CdS buffer layer, acting like a n-type dopant and decreasing the net acceptor concentration in the chalcopyrite phase.[42] However, since the absorber/CdS stack is obtained from the etched finished solar cells (Experimental methods), this is not the reason for the higher non-radiative loss in the cell. Multiple measurements are performed to correctly evaluate the optical loss. This loss, being in the range of the QFLS variation within the sample, is reported with a relevant error bar (Figure 4c). Urbach energies are not substantially changing from the absorber when the top layers are deposited, even though are slightly higher for the final cell, in accordance with what was previously noted in Figure 3a and Figure 3b. There is a small difference between the tail states calculated in Figure 3a on the absorber/CdS stack, which are slightly higher. The absorptance extracted from the large spot measurements on absorber/CdS stacks and on the cells is shown in **Figure 4b**. The finished device exhibits a very slight reduction in A(E) between 0.95 and 1.05 eV, arising presumably from the parasitic absorption of the TCO.[12] It should be noted, that in both cases, the analysis fixes the absorptance =1 at high energies. In **Figure 4c**, we display all the loss mechanisms, leading to the finished cell $V_{OC}$. Non-radiative losses in the absorber are the main fundamental loss, accounting for more than 150 mV loss in the finished cell. Improvements in the optoelectronic quality of the bulk of the chalcopyrite are essential to reduce this loss. This has been partly achieved by optimized alkali supply, which improved the $V_{OC}$ by 15meV.[21] Further passivating the front surface has recently lowered non-radiative losses in novel devices.[8,9] Radiative losses are calculated from σ with a formerly reported formula[25,26,43] (Equation 9). Generation losses, arising from the $J_{SC}$ loss (light grey in Figure 1c), contribute only to a small extent. The total sum of the various loss processes in Figure 4c are compatible with the $V_{OC}$ deficit of the finished device (brown bar), which is determined as the difference between the SQ-$V_{OC}$ for 1.00 eV band gap[16] and the actual $V_{OC}$.

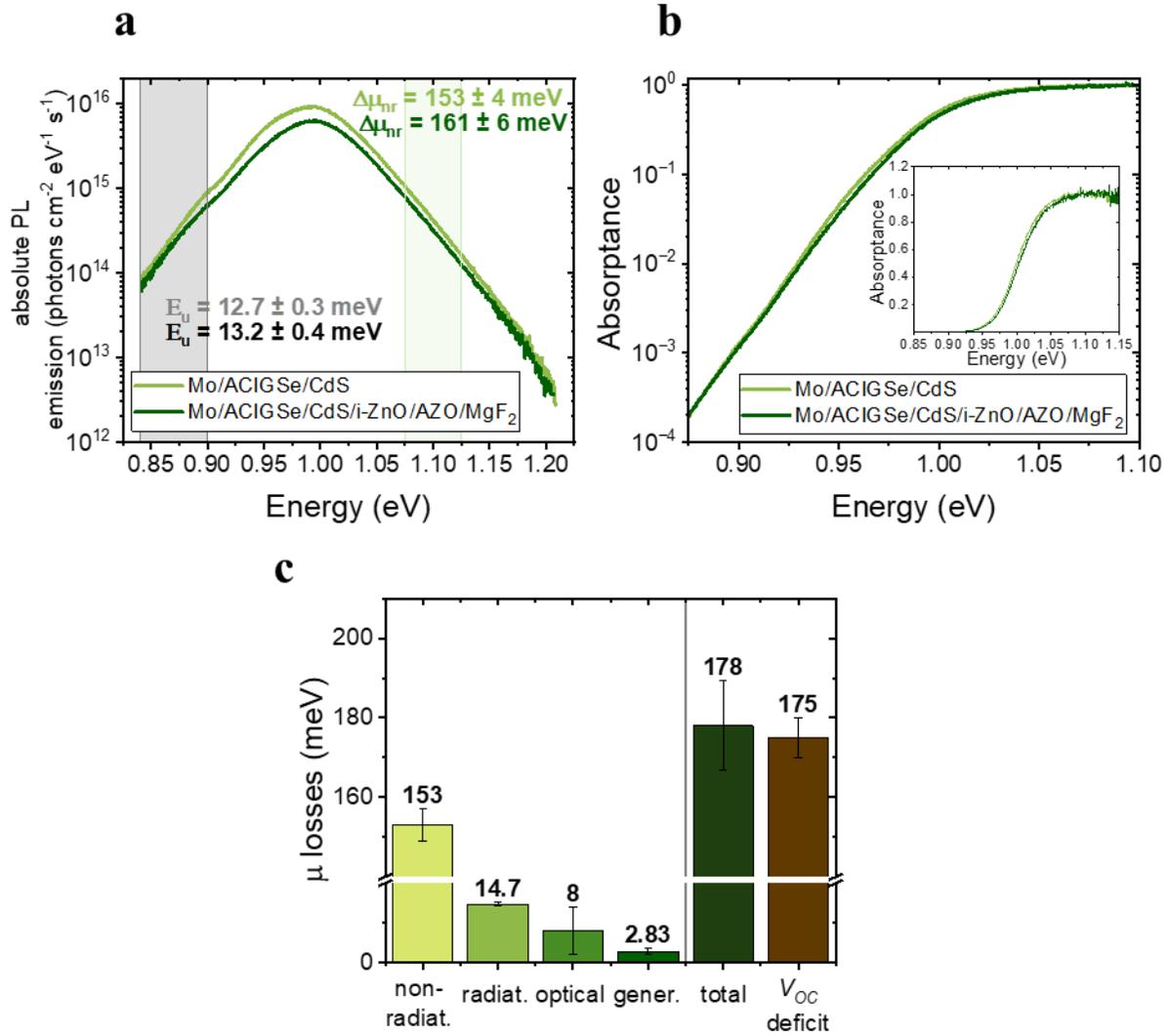

**Figure 4.** a) absolute PL spectra (in both cases, grids were on top of the samples) of the absorber/CdS stack (light green, Absorber/CdS spot b3 from Table 3) and the finished cell (dark green, Finished device spot b3 from Table 3) measured with a 5 mm radius 637 nm laser (in green the QFLS and in grey the Urbach energy fit areas, extracted with Equation 3 and Equation 7, respectively); b) log plot of the PL extracted absorptance spectra (inset linear scale); c) various losses mechanisms in the $V_{OC}$.

Figure 5 displays the onset of the EQE spectrum of cell B (Figure 1c), alongside the rescaled absorptance spectrum extracted for PL measurement on the finished cell spot b3 (Figure 4b and Table 3). The absorptance spectrum from PL is not absolute, since we assume A=1 in the PL analysis, because no reflectance spectrum has been measured with a spectrophotometer. The reflectance would be required for an absolute PL absorptance spectrum.[18] The finished cell has grids on top, which prevent obtaining the correct reflectance with a spectrophotometer. In fact, the EQE measurement (Experimental methods) is performed on the active area only, the cell is excited with a light beam smaller than the distance of the Ni/Al grids. Therefore, the PL absorptance spectrum is scaled to agree with EQE spectrum in the energy range, where both are reliable and not noisy from 0.95 to 1.05 eV. Minimal differences are noted between the two spectra. In the range between 1.05 and 1.10 eV, the PL absorptance is slightly higher than the EQE, possibly indicating a very small collection loss. However, the minor variation between the two spectra indicates that collection losses do not play a notable role in these high efficiency bottom cells.

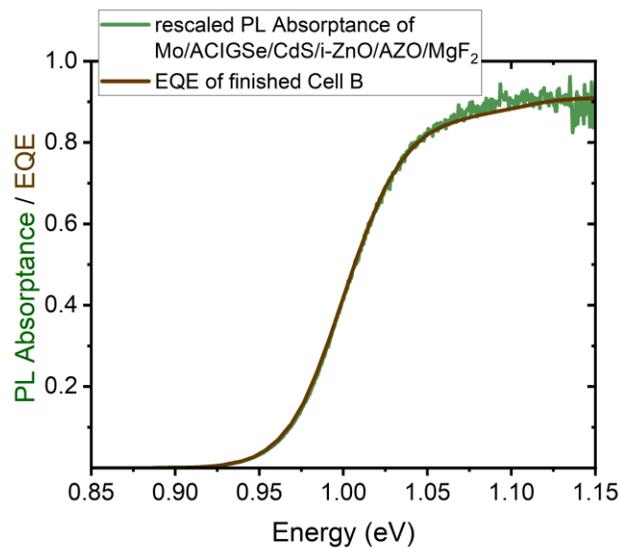

**Figure 5.** a) rescaled absorptance spectrum extracted from the PL on the finished cell spot b3 (Table 3 and Figure 4b); b) EQE spectrum of the cell B.

Fill factor (*FF*) is the third parameter defining the PCE of a solar cell. With a given $J_{SC}$ and $V_{OC}$, *FF* is determined by shunt resistance, series resistance and the diode factor (or ideality factor). Since we determine the diode factor from a fit to the *J-V* characteristics, from the PL or the electroluminescence (EL) power law, we differentiate between the electrical (EDF), optical (ODF) and EL diode factor (ELDF), respectively.[19,20,32] Shunt (> $2 \cdot 10^3$ Wcm$^2$) and series resistances (< 0.5 Wcm$^2$) appear not to be limiting elements of *FF*, as seen in the one-diode fits in **Figure 6a** and **Figure 6b**. Therefore, we investigate the EDF, which results in particularly high values from the fit of *J-V* curves (Figure 6a, Figure 6b). The ODF is determined via the emitted photon flux from the absorber/CdS stack or finished cell, varying the generation flux: the slope of a linear fit in the log-log plot gives the ODF (**Equation 11**). The ODF is the lower limit of the EDF.[19,32] The two are the same if the same recombination mechanisms are present in the absorber alone compared with the complete device.[19,20] However, we need to remark that, due to the low n-doping of the CdS buffer layer, there is no space charge region (SCR) in the absorber/CdS stack. Therefore, the ODF is measured in the neutral region where only the minority carrier quasi-Fermi level shifts. This differs from EDF, where the presence of the TCO creates the SCR where both quasi-Fermi levels can shift. In the case of predominant recombination in the SCR, the ODF on the absorber and the EDF on the cell may differ, with the EDF being higher than the ODF. In addition to EDF from *J-V* scans, the ODF on the absorber/CdS stack and the finished cell, we measure the electroluminescence diode factor (ELDF). For the ELDF, the emitted photon flux from the cell is measured varying the concentration of electrical injected charge carrier.

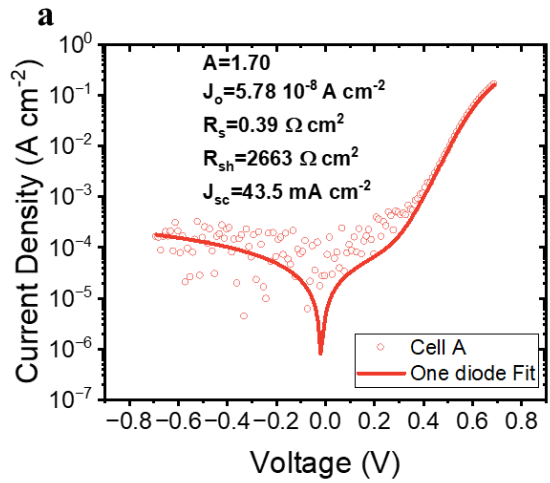
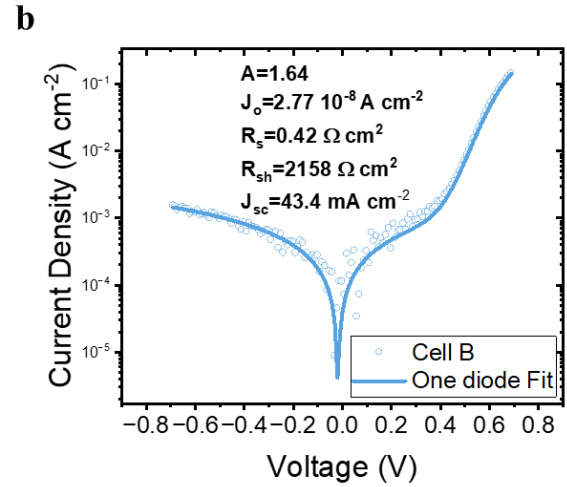
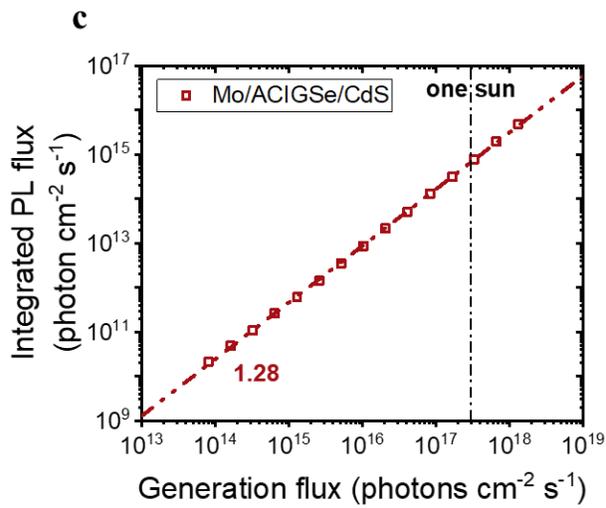
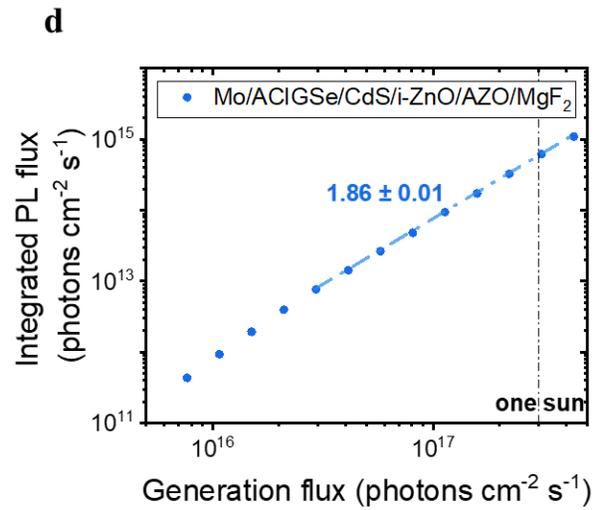
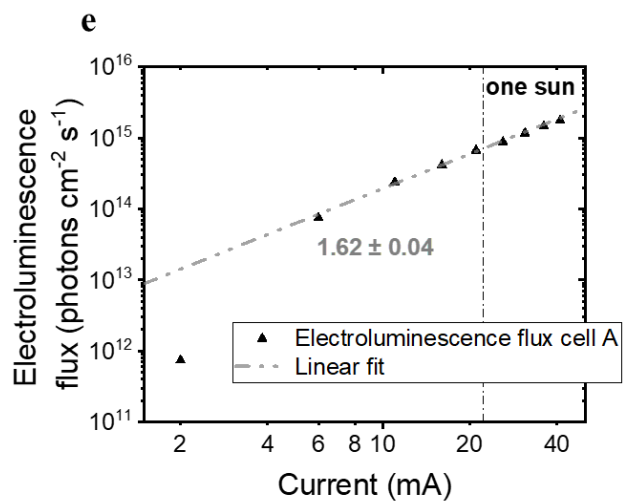

**Figure 6**. a, b) one-diode fit of the *J-V* dependence under illumination (shifted up by $J_{SC}$) of cell A (a) and cell B (b) (for linear plots see Figure 1a); c) generation flux-dependent integrated PL flux of the absorber/CdS stack (fit error below 0.01); d) generation flux-dependent integrated PL flux of the finished cell area b (grey area in Figure 8); e) injected carrier-dependent integrated EL flux of the finished cell A. Linear fits in the Figure 6c, Figure 6d and Figure 6e are performed in a log-log plot, as the ODF follows a power law (Equation 11).

In **Figure 6c**, we plot the emitted photon flux from the absorber/CdS stack with varying generation flux. We show the linear fit of the log plot allowing us to extract the ODF. It is notable that the absorber has already an ODF above 1, although the absorber is p-type and the illumination is in the low excitation regime. This is likely due to metastable defects inside the absorbers, changing from donor to acceptor behavior upon illumination.[6,32,44] The ODF of the absorber film of 1.28 is lower than the 1.40 previously reported for high efficiency bottom cell without Ag[32] (dashed line in **Figure 7**). It was observed before that addition of Ag lowers the ODF.[32] There is indication that the activation energies for the transformation between the donor and acceptor state are higher when Ag is added.[32] The electroluminescence diode factor [45,46] (ELDF, **Figure 6e**), obtained from varying the injected current, is very similar to EDF value on the same cell A, as expected. We need to consider that the collection spot of the luminescence in the ELDF is much smaller than the full cell, so slight inhomogeneities may explain the small difference. Both EDF and ELDF display the diode factor of the solar cell, which is higher than the diode factor of the absorber/CdS stack, because the cell has an additional recombination channel in the SCR which is not present in the absorber/CdS stack. In addition, we perform ODF measurement on the finished cell (**Figure 6d**). The reported value (1.86) is extracted from the linear fit in a log-log plot only at higher excitation. Unexpectedly, the ODF of the cell is higher than the EDF of the same cell. We offer three explanations: (i) the excitation of the ODF measurement is with a red laser, i.e. is missing the longer wavelength light which is absorbed outside the SCR (where the diode factor is close to 2) in the neutral region (where the diode factor is given by the ODF of the absorber, i.e. 1.3). However, this difference in generation location would only make a difference in the diode factor if diffusion was weak. Otherwise, the carriers diffuse throughout the absorber thickness. Since we do not see any indication of collection losses, the difference in excitation light may not be explanation. (ii) both the ODF on the cell and the ELDF on the cell tend to deviate from the straight line at lower excitation/injection. This effect is attributed to the voltage dependence of the SCR extension (at lower voltages the SCR is large and thus the over-all diode factor closer to 2) The EDF and the ELDF are measured near and above 1 sun $V_{OC}$ (see the 1 sun line), whereas the ODF on the cell is measured mostly below 1 sun $V_{OC}$ (due to the limitations in laser power), i.e. in a region where the SCR is larger and therefore the diode factor closer to 2. The shunt resistance can also "bend" diode factor measurements at low voltages or low current injection. The shunt resistance is likely the reason of the deviation of the first point in the ELDF measurement at 2 mA. However, in the ODF measurement, that shunt does not play a role, since the cell is always at $V_{OC}$ and no current in flowing. (iii) red illumination leads to a different band bending in the absorber and in the top layers,[22] compared to white illumination. Moreover, the lack of blue illumination causes a formation of a p+ layer close to the interface. Under red light, interface selenium-copper divacancy acts as shallow acceptors and interface $In_{Cu}$ acts as deep neutral defects. Only holes, minority carriers at the interface, generated by blue photons in the buffer relax these defects occupied by electrons.[47] These effects may change the dominant recombination region and influence the diode factor.

All diode factors, extracted by our different measurements, are displayed in Figure 7 and compared with literature values. The first chart box reports all the formerly measured EDFs of the cells on the sample NaF10.[21] They show quite a variety. The EDFs obtained from our $J$-$V$ scans in Figure 6a and Figure 6b (black diamonds) and the ELDF of Figure 6e (empty triangle) are compatible with them. The ODF on the finished cell of Figure 6e (empty circle) is higher than the average, as discussed above. The ODF of the absorber/CdS from Figure 6c is considerably lower than any of the EDFs, indicating again that the cells exhibit an additional recombination mechanism, namely SCR recombination, which is not present in the absorber alone. An increment from the absorber ODF to the EDF value is a trend that we see also in other chalcopyrite bottom cells. We present the results of pure CISe cells without Ga in the absorber passivated on the back side with a hole transport layer (HTL).[48] The final EDFs of these HTL samples are comparable with the ones in this study. However, the absorber ODF is larger, supposedly due to the lack of Ag alloying. The ODF of these pure CISe samples is close to the ODF of earlier low-gap CIGSe absorbers from EMPA (without Ag),[32] as depicted by the dashed line in Figure 7. In the same graph, EDFs from the recent record CIGS bottom cells from Wuhan university are shown.[8] The series without Ga gradient in the front side has somewhat lower EDFs, but still higher than the ODFs of CI(G)Se absorbers without Ag (dashed line and red open square in Figure 7). However, when Ga is also introduced at the front interface, EDFs are reducing to values similar to absorber ODFs without front Ga gradient and without Ag. This trend is also found in CIGS bottom cells on flexible substrates.[9] These observations support the model that the EDF of the low gap CIGSe cells is high because of SCR recombination. Adding Ga near the front, i.e. in the SCR, increases the band gap and therefore reduces recombination in the SCR. We may exclude interface recombination as the cause increasing the diode factor. In fact, interface recombination leads to a EDF equal to 1 in the finished device.[49] Only the hole quasi fermi level moves with voltage, due to their low concentration at the interface induced by band bending. Furthermore, band bending is present in the cell, compared to the flat bands of the absorber/CdS stack. Thus, interface recombination would be more important in the absorber/CdS stack than in the cell. So, the model that emerges from this comparison is: pure CuInSe$_2$ contains SRH recombination centres, which lead to SCR recombination which increases the EDF and reduces the *FF*. This model explains the difference between the absorber ODF and the device EDF. Adding a Ga front gradient reduces SCR recombination due to the higher band gap, thereby lowering the EDF. It should be noted that a higher Ga content changes the metastability behavior.[32,50] Further research should be carried out to better understand which defects cause the SCR recombination (could be the well-known defect at 0.8 eV[51]) and how to reduce their presence without using additional Ga at the front. A low bandgap CuInSe$_2$ cell without any Ga gradient would be ideal to allow for thinner absorbers and to reduce radiative losses.[18]

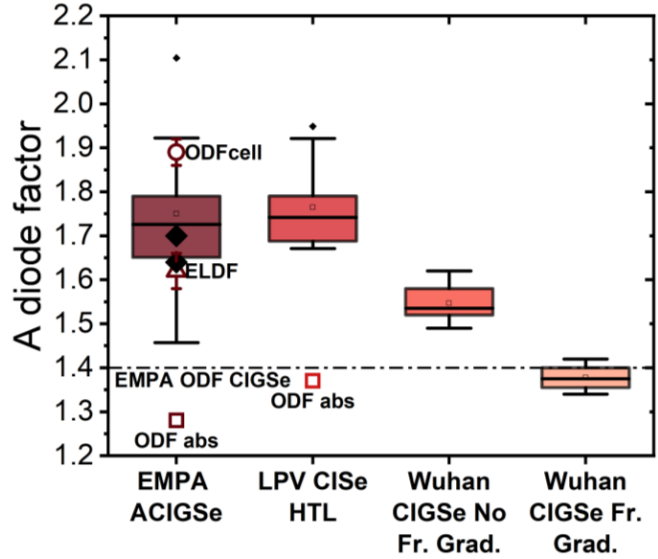

**Figure 7.** Comparison of electrical diode factors with literature; in dark red the box chart with the EDFs of the herein studied ACIGSe sample, as previously reported.[21] The EDF of the two cells in this study are displayed as black diamonds (data from Figure 6a and Figure 6b), ODF of absorber/CdS stack in empty square (data from Figure 6c), ELDF in empty triangle (data from Figure 6e) and ODF of finished cell in empty circle (data from Figure 6d). In red, the box plot of EDFs from HTL passivated CISe cells made at LPV, with their absorber ODF.[48] In light red, the EDFs of the Wuhan CIGSe cells with and without Ga front gradient.[8] The dashed line represents the absorber ODF on a previously published EMPA CIGSe (without Ag) cell.[32]

## 3. Conclusion

In conclusion, we analyze the fundamental losses in state-of-the-art bottom cells for tandem applications. Our results highlight the non-radiative loss in the bulk of the absorber as the main $V_{OC}$ loss. They account for more than 150 mV, while radiative and optical losses amount to 15 and 8 mV, respectively. The generation loss (or short circuit loss) of $V_{OC}$ is negligible, as only about 3 mV are lost. Interestingly, the absorption edge broadening and tail states of these bottom cells are already determined by the absorber properties themselves and the addition of the top layers does not play a role. Total $J_{SC}$ loss amounts to about 5 mA cm$^{-2}$, and is due in the short wavelength region to parasitic absorption in the buffer and window layers and in the long wavelength region to absorption losses in the absorber itself. The *FF* is mainly limited by a large diode factor. The EDF is greater than the absorber ODF, due to additional SCR recombination when p-n junction and band bending are created. This trend is commonly observed in samples, where the SCR is pure low-gap CuInSe$_2$.

Finally, we consider possible improvements and their effect on solar cells efficiency. Assuming the absorber shows 1% PLQY, as the current 1.13 eV bandgap ACIGSe record,[13] we would reduce non-radiative losses at 296K to 117 meV. Taking into account the same amount of radiative losses of the device here analyzed (15 meV), the resulting $V_{OC}$ would be 634 mV, improving the efficiency by 1.2% absolute. This is comparable to the $V_{OC}$ of the Wuhan 1.01 eV bandgap CIGSe record solar cell,[8] when a Ga front gradient is introduced. However, the latter is detrimental to the current, so new strategies must be implemented. The introduction of an alternative buffer without parasitic absorption, such as Zn(O,S)[14] or ZnSnO[52,53] buffer layers, and of a TCO with less free carrier absorption, such as bias ZnO[24] may halve the current loss and allow a $J_{SC}$ of 45.9 mA cm$^{-2}$, which would improve the efficiency by 0.2% absolute. If the EDF would be equal to lowest ODF measured on chalcopyrite materials (1.1),[32] $FF$ could be improved to 79.3% and the efficiency higher by 1.9% absolute. In this scenario, the final efficiency of the enhanced device would be 22.8%. Our detailed loss analysis has identified high diode factor due to SCR recombination and non-radiative absorber recombination leading to $V_{OC}$ loss as the main loss mechanism in these low gap cells.

## 4. Experimental methods

### 4.1. Sample preparation

The solar cells are structured as Glass/Mo/ACIGSe/CdS/i-ZnO/AZO/Ni-Al grids/MgF$_2$. Their preparation is described as sample NaF10 in a previous paper.[21] The cells measured, with an area around 0.5 cm$^{-2}$, are obtained from the same absorber. In order to analyze the absorber/CdS stack, the finished cells are cut from the area abc (Figure 8) and etched for 30 seconds in a 10% acetic acid DI water solution, to remove the i-ZnO/AZO/MgF$_2$ layers, but not the CdS (yellow area in Figure 8).[54] The grids are not removed by the etching. A part of the abc area is left with the top layers to perform optical measurements on the finished cells (grey area in Figure 8).

### 4.2. Absolute photoluminescence

Absolute PL was measured using in-house spectral and intensity calibration of the PL spectra. The absorber/CdS stack was excited with a continuous-wave laser (660 nm) of 2.2 mm diameter at a photon flux density equivalent to 1 sun. While the finished cells (cut from the abc area as shown in dark grey in Figure 8) are excited with a continuous-wave laser (637 nm) passing through a beam-expander to obtain a beam of 5 mm diameter, with a photon flux density equivalent to 1 sun. The beam diameter is larger than the analyzed abc area of the finished cells (dark grey are in Figure 8). For the finished devices, a larger laser beam spot, covering the full device, is needed to correctly determine the open circuit QFLS. Indeed, the junction of the cell drives a current and the TCO is a nearly equipotential surface. Therefore, an electrical connection is present between the different parts of the cell. If this is illuminated with a small spot, the electrical connection exists between a dark (0 V) and an illuminated part. The latter should be at open-circuit condition, but it cannot be due to the current flowing from the illuminated to the dark part. This makes necessary to shine the laser beam on the whole devices. Since the collection spot is much smaller, around 0.3 mm, we still measure several spots on each cell area. Measurements on the absorber/CdS stack from the two different setups give compatible QFLSs (Tables 1 and 3).

The PL signal was collected and led by two parabolic mirrors, a 550 µm optical fiber to a monochromator and then an InGaAs array detector. The InGaAs array detector used for the larger beam spot has higher dark current and the measurements results are noisier. This prevents to analyze the absorptance broadening of the spectra obtained with the larger beam spot.

The quasi-Fermi level splittings were extracted by fitting the high-energy tail of the PL spectrum with Equation 3,[29,33] assuming an absorptance $A(E) = 1$. This assumption underestimates the QFLS by about 5 meV, as $A(E) < 1$ in this region.[18] Our fit is based on Planck's generalized law (Equation 2) under the Boltzmann approximation at a fixed temperature of 296 K, which is the measured temperature in the laboratory.

The different names of the spots are referred to the different cut and etched cells of the same absorber (Figure 8). On the etched cells and finished cells, multiple measurements are taken, so numbers are added. Even in the case of the larger beam, multiple spots are measured, as the collection area is much smaller than the illuminated area. The two finished cell A and B, used for the electrical measurements, belong to the same sample, as shown in Figure 8. The sample is very homogeneous, as seen in the parameters of all the NaF10 cells in the previous paper.[21]

Furthermore, different calibrations and alignments of PL/EL setup introduce systematic errors of about 5 meV. This must be taken into account when comparing results, for example, between measurements using large and small laser beam spots. We also remark that temperatures of the optical and electrical measurements are slightly different (296K for PL and 298K for $J$-$V$) and this may create a difference of about 3-4 m(e)V.

Equation 5, used to fit the derivative of the PL absorptance or EQE, contains a constant term $C_0$ because the absorptance does not reach exactly one. This term is necessary to have a correct evaluation of the broadening. However, it results in a very low value.

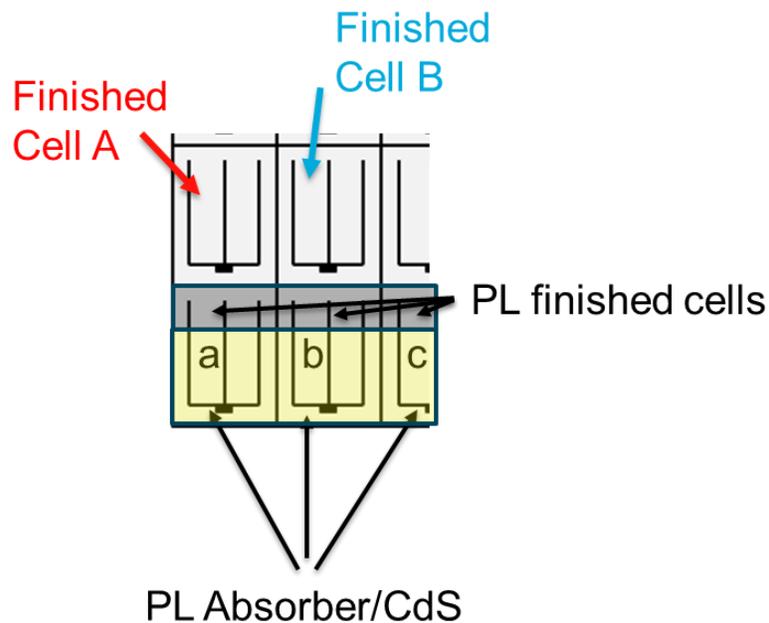

**Figure 8**. Detail on the sample analyzed in this study.

### 4.3. External quantum efficiency

A homebuilt setup was used to determine EQE spectra of the cells with a four-probes configuration. This setup has a lock-in amplifier to measure the photocurrent of the cells under a chopped illumination. The light source is a halogen or a xenon lamp passing through a monochromator. The spectral intensity of the illumination is measured by calibrated Si and InGaAs detectors.

### 4.4. *J-V*

*J-V* curves were performed using a class-AAA AM 1.5 solar simulator at a temperature of 298K, with a four-probe configuration. The one diode fits, based on the Orthogonal Distance Regression (ODR),[55] are obtained by a home-made script.

### 4.5. Electroluminescence measurements

The EL measurements were performed on solar cells using a four-probe contact configuration. We injected a current into the solar cell. The collection setup is the same used for PL with larger beam spot. Therefore, the EL emission was collected with two off-axis parabolic mirrors and detected by an InGaAs detector.

### 5. Appendix

Figure 9 displays all the analysis on the measured absorber/CdS spots. Results are reported in Table 1. The sample is homogeneous and no essential differences are noted between the different spots. Measurements on spot a1 and b1 are noisier, due a noisier calibration correction. However, they are coherent with the rest of the spots.

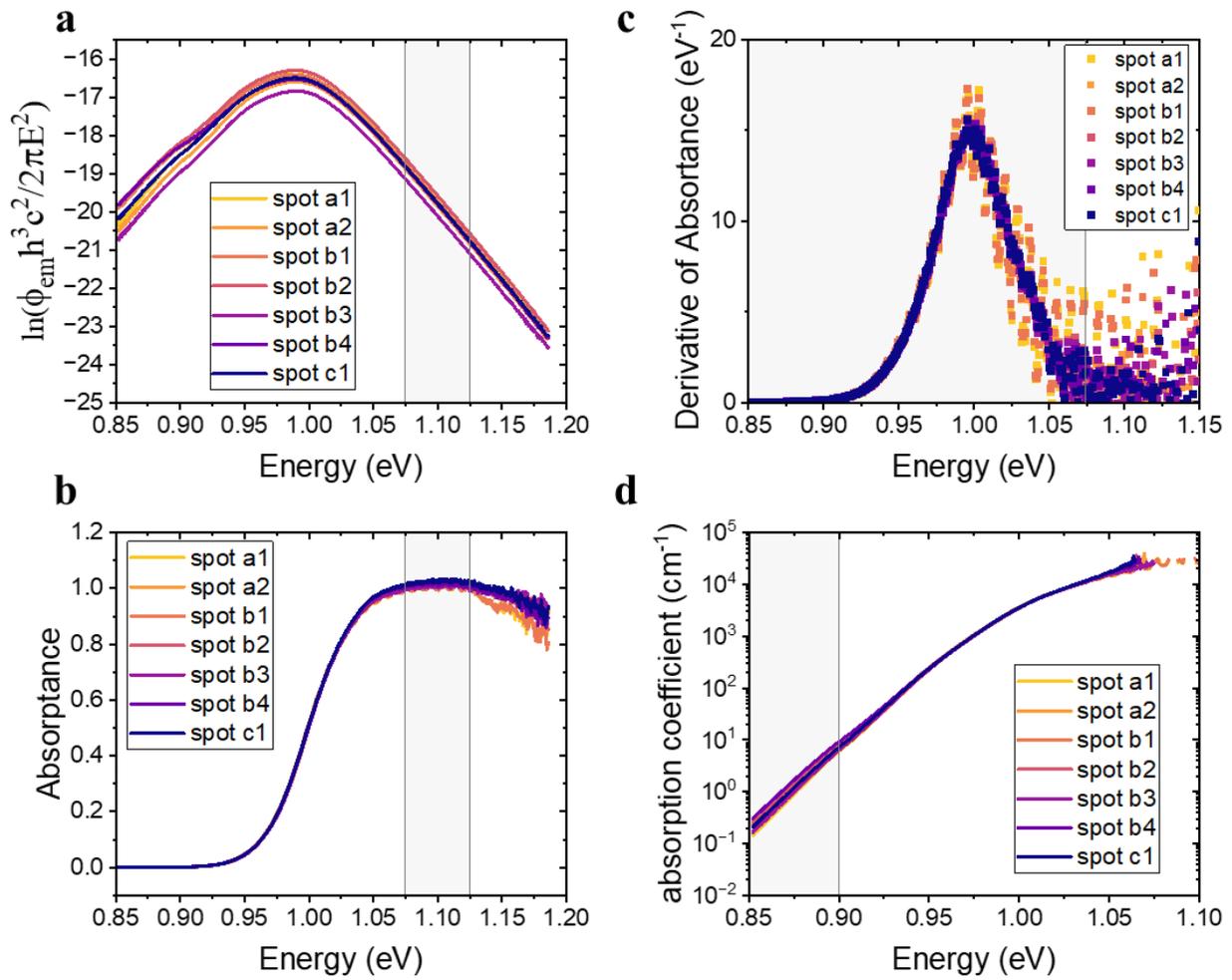

**Figure 9.** a) absolute PL spectra of all the absorber/CdS stack spots; b) their extracted absorptance spectra; c) first derivative of the extracted absorptance spectra; d) absorption coefficient from the PL extracted absorptance spectra with its Urbach energy fit. Fitting ranges in grey.

Figure 10a shows all the PL measurements performed with larger beam on absorber/CdS stack (light green) and finished cell (dark green). Their PL absorptance spectra in Fig. 10b confirm that there is not variation in the absorption edge between the spots taken on the absorber/CdS stack, as well as the ones on the finished devices. QFLSs, non-radiative losses and Urbach energies of all the spots are documented in Table 3.

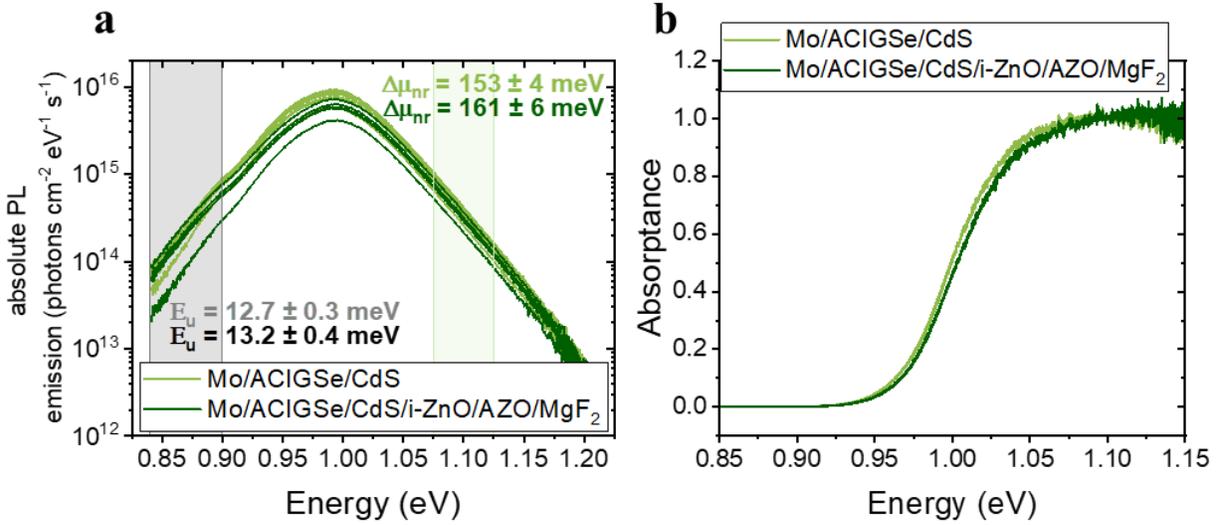

**Figure 10.** a) absolute PL spectra (in both cases, grids were on top of the samples) of the absorber/CdS stack (light green) and the finished cell (dark green) of all the spots analyzed here with the large beam spot (in green the QFLS and in grey the Urbach energy fit areas); b) plot of their PL extracted absorptance spectra. No differences are noted between the different spots.

**Table 4.** List of symbols.

| Symbol | Meaning (unit) | Symbol | Meaning (unit) |
|---|---|---|---|
| $E$ | Photon Energy (eV) | $\Phi_{BB}(E)$ | Black body emission spectrum (1 · cm$^{-2}$s$^{-1}$V$^{-1}$) |
| $E_g$ | Band gap energy (eV) | $J_0^{SQ}$ | Shockley-Queisser radiative saturation current density (A · cm$^{-2}$) |
| $F_{Gen}^{Sun}$ | Sun generation Flux (cm$^{-2}$s$^{-1}$) | $J_{SC}$ | Short circuit current density (A · cm$^{-2}$) |
| $F_{Gen}^{Laser}$ | Laser generation Flux (cm$^{-2}$s$^{-1}$) | $J_{SC}^{SQ}$ | Shockley-Queisser short circuit current density (A · cm$^{-2}$) |
| $E_u$ | Urbach energy (eV) | $Y_{PL}$ | PL quantum yield |
| $A(E)$ | Absorptance spectrum from PL | $Y_{EL}$ | EL quantum yield |
| $QE(E)$ | External Quantum efficiency spectrum | $V_{OC}$ | Open circuit voltage (V) |
| $T$ | Temperature (K) | $\Delta\mu$ | Quasi Fermi level splitting (QFLS) (eV) |
| $G$ | Generation flux (cm$^{-2}$s$^{-1}$) | $V_{OC}^{SQ}$ | Shockley-Queisser open circuit voltage limit (V) |
| $\alpha$ | Absorption coefficient (cm$^{-1}$) | $\delta\Delta\mu^{rad}$ | Radiative QFLS loss (eV) |
| $d$ | Sample thickness | $\delta\Delta\mu^{sc}$ | Short-circuit QFLS loss (eV) |
| $\sigma$ | Absorptance edge broadening parameter (eV) | $\delta\Delta\mu^{nr}$ | Non-radiative QFLS loss (eV) |
| $n_{opt}$ | Optical diode factor | $V_{OC}$ | Open-circuit voltage (V) |

## Data availability

All data supporting this study are available in Zenodo.[56]

# Notes

The authors declare no competing financial interest.

# Author Contributions

F.L. conceived the idea and designed the experiments, measured the samples by PL and EL, conducted the *J-V* and EQE measurements, and wrote the first draft of the manuscript. S.G. contributed to the development of the EL setup and the data analysis. M.K., S.N. and R.C. provided the $(Ag,Cu)(In,Ga)Se_2$ samples, with their additional data. S.S defined the project, conceived the idea, acquired the funding, guided the data analysis and discussion and contributed to the writing of the manuscript. All the authors contributed to results discussion and to revise the paper.

# Acknowledgments

Financial support from the Luxembourgish Fonds National de la Recherche (FNR) in the framework of the project FULL CORE/22/MS/17138895/FULL. For the purpose of open access, the author has applied a Creative Commons Attribution 4.0 International (CC BY 4.0) license to any Author Accepted Manuscript version arising from this submission.